# A review of the scattering parameter extraction method with clarification of ambiguity issues in relation to metamaterial homogenization

S. Arslanagić [1], T. V. Hansen [1], N. A. Mortensen [2], A. H. Gregersen [1], O. Sigmund [3], R. W. Ziolkowski [4], and O. Breinbjerg [1]

[1] Department of Electrical Engineering, Technical University of Denmark, Build. 348, Ørsteds Plads, DK-2800 Kgs. Lyngby, Denmark

[2] Department of Photonics Engineering, Technical University of Denmark, Build. 343, Ørsteds Plads, DK-2800, Kgs. Lyngby, Denmark

[3] Department of Mechanical Engineering, Technical University of Denmark, Nils Koppels Allé, Build. 403, DK-2800, Kgs. Lyngby, Denmark

[4] Department of Electrical and Computer Engineering, University of Arizona, 1230 E. Speedway Blvd., Tucson, Arizona, 85721-0104, USA

**Abstract:** The scattering parameter extraction method of metamaterial homogenization is reviewed to show that the only ambiguity is the one related to the choice of the branch of the complex logarithmic function (or the complex inverse cosine function), whereas it has no ambiguity for the sign of the wave number and intrinsic impedance. While the method indeed yields two signs of the intrinsic impedance, and thus the wave number, the signs are dependent, and moreover, both sign combinations lead to the same permittivity and permeability, and are thus permissible. This observation is in distinct contrast to a number of statements in the literature where the correct sign of the intrinsic impedance and wave number, resulting from the scattering parameter method, is chosen by imposing additional physical requirements such as passivity. The scattering parameter method is reviewed through an investigation of a uniform plane wave normally incident on a planar slab in free-space, and the severity of the branch ambiguity is illustrated through simulations of a known metamaterial realization. Several approaches for proper branch selection are reviewed and their suitability to metamaterial samples is discussed.

## 1. Introduction

Since the pioneering work of Veselago [1], where the first systematic study of materials with negative material parameters was performed, and following the initial realizations of these and related structures some 30 years later [2-4], a tremendous amount of work describing and demonstrating their interesting perspectives emerged [5-7]. Materials characterized by a negative permittivity and permeability belong to the broad class of artificially constructed materials, termed metamaterials (MTMs). MTMs are highly inhomogeneous structures composed of periodic or random arrangements of scattering elements inside a host medium, and they possess properties generally not found in natural materials.

With numerous indicators of the huge potential of MTMs, *e.g.*, providing alternative routes to miniaturization of a number of electromagnetic devices [5-7], and facilitating perfect lenses [8], and cloaks [9], there continues to be a need for proper characterization of MTMs in order to better understand and further exploit their properties. In this regard it is useful to recall that our understanding of electromagnetic wave interaction with ordinary materials, being inhomogeneous at the atomic scale, is facilitated through the introduction of material parameters such as permittivity and permeability. The associated homogenization process, i.e., the appropriate averaging that provides the material parameters, is enabled by the fact that the electromagnetic response is due to a large ensemble of atoms with extent and separation distances far below the operating wavelength, rather than due to the individual atoms constituting the material. It has been proposed that a MTM composed of resonant scattering elements whose size and spacing are far below the operating wavelength should respond to electromagnetic waves in a similar (ideally identical) way and, consequently, can be characterized by effective material parameters. Several homogenization approaches to accomplish the task have been proposed in the literature. They include a variety of field-averaging approaches [10-15], the curve-fitting approach [16], dispersion equation method [17], and the scattering (S) parameter extraction method [18-23]. The latter has become a prime tool for MTM characterization. This technique facilitates the extraction of the permittivity and permeability from the measured – or otherwise known – S-parameters, with the wave number and intrinsic impedance obtained as intermediate steps. The S-parameter method applied in [18-23] to MTM samples illuminated by normally incident plane waves is broadly known as the Nicolson-Ross-Weir method. This method has been used for experimental characterization of many homogeneous materials [24-26]. The method has moreover been extended in [27] for characterization of MTMs in the case of obliquely incident plane waves – an issue of prime importance in cases where anisotropy or spatial dispersion cannot be neglected. Despite its widespread use for MTM characterization, the S-parameter extraction method is ambiguous and does not readily give a unique value of wave number and, thus, permittivity and permeability. The ambiguity can be explained in terms of the associated Bloch state physics, see, e.g., [28]. It appears mathematically as branches of the complex logarithmic function. Henceforth, this ambiguity is referred to as the branch ambiguity. Moreover, a number of works [18-23] have noted the inability of the S-parameter method to provide a unique sign of the intrinsic impedance and wave number (or refractive index). These claims arise because the method directly gives both signs. This ambiguity, henceforth referred to as the sign ambiguity, was resolved by use of additional physical arguments such as passivity.

In addition to these ambiguities, a few other challenges are generally associated with the common S-parameter method. These include the occurrence of non-physical phenomena occurring near the Fabry-Pérot resonances of the MTM sample, *i.e.*, when the sample thickness is an integer multiple of half of the wavelength inside the sample. These phenomena, which are introduced in the method through the intrinsic impedance, are often associated with numerical and experimental noise [29, 30]. Several methods have been proposed for their compensation [30]-[32]. Yet a drawback of the common S-parameter method, which assumes well-defined sample boundaries, is its inability to properly account for the physical boundaries of a realistic MTM sample and its length. To properly account for these boundary effects, significant efforts have been reported on the use of transition layers [33, 34], as well as the so-called generalized sheet transition conditions [35, 36], as a means of augmenting the method to provide more accurate extracted MTM parameters.

The purpose of the present work is to review the S-parameter extraction method and to clarify its ambiguity issues. We compare different – but equivalent – formulations of the method; and we show that it possesses only one ambiguity, namely the branch ambiguity, which appears in the real part of the wave number. We focus on a case of a normally incident plane wave on a planar homogeneous slab in free space. In contrast to previous reports [18-23] we find that there is no sign ambiguity for the wave number and intrinsic impedance since both signs lead to the same permittivity and permeability and neither can thus be discarded by physical arguments such as passivity. The simple reason is that the wave number and intrinsic impedance – unlike the permittivity and permeability – are not fundamental quantities in Maxwell's equations or its constitutive relations, but are derived quantities that are introduced for convenience. When introduced they can be defined with one sign or the other. As long as either definition is followed stringently, the initial choice of sign remains valid. As to the branch ambiguity, we demonstrate that it is a consequence of a particular set of conditions. We review various approaches to resolve this ambiguity and discuss their MTM applications.

Though outside the main scope of this manuscript, it is pertinent to address briefly the issue of how the permittivity and permeability determined from the S-parameter extraction method can be interpreted for MTMs. Of course, the determined permittivity and permeability are *equivalent* material parameters in the sense that a slab of a homogeneous material with the determined permittivity and permeability will give the same S-parameters as the actual MTM slab. However, this does not necessarily imply that the determined permittivity and permeability are also *effective* material parameters in the sense that there is a macroscopic field inside the MTM slab similar to the field inside the homogeneous slab. Any MTM slab – no matter the coarseness of its structure – can be attributed a set of equivalent material parameters. However, in order for these to be also effective material parameters when structural periodicity is involved, it is also necessary for the periodicity to be very small compared to the wavelength. For the purpose of the present work it is not necessary to distinguish between equivalent and effective material parameters. In the following we simply refer to permittivity, permeability or material parameters. The distinction between equivalent and effective material parameters has been made in [37].

The manuscript is organized as follows. In Section 2, we recall the solution to the forward problem of a normally incident uniform plane wave on a planar slab in free space. We demonstrate that the wave number and intrinsic impedance can be introduced with either sign without changing the physics of the problem. A number of different – but equivalent – expressions for the S-parameters are reviewed and we identify the different sets of permittivity and permeability that lead to the same S-parameters. These constitute the ambiguous solutions for the S-parameter extraction method. In section 3 we discuss the inverse problem of determining the permittivity and permeability from the S-parameters using the expressions established in Section 2. We demonstrate that both signs of the wave number and intrinsic impedance that follow from the inversion of the S-parameter expressions are equally valid and that the only ambiguity is the branch ambiguity related to the branch of the complex logarithm (or complex inverse cosine). Section 4 illustrates the significance of the choice of the branch through simulations of a slab with constant material parameters and a specific MTM design [38]. Section 5 provides a review of the potential approaches for solving the branch ambiguity and discusses their MTM applications. Finally, section 6 summarizes and concludes this work. Throughout the manuscript, the time factor $\exp(j\omega t)$, with $\omega$ being the angular frequency, and $t$ being the time, is employed and suppressed.

## 2. Forward problem

### 2.1 Configuration

Let a uniform plane wave be normally incident upon a planar slab of a simple magnetodielectric material, see Figure 1. The slab is located in free-space with permittivity, $\varepsilon_0$, permeability, $\mu_0$, and thus, the wave number, $k_0 = \omega\sqrt{\varepsilon_0\mu_0} > 0$, and intrinsic impedance $\eta_0 = \sqrt{\mu_0/\varepsilon_0} > 0$. It has a thickness $d$ and consists of a simple material characterized by a permittivity and a permeability, denoted by $\varepsilon_s(\omega) = \varepsilon_s' + j\varepsilon_s''$ and $\mu_s(\omega) = \mu_s' + j\mu_s''$, respectively, leading to a wave number

$k_s(\omega) = k_s' + jk_s''$ and an intrinsic impedance $\eta_s(\omega) = \eta_s' + j\eta_s''$. It is well-known that in a passive material, $\varepsilon_s'' \leq 0$ and $\mu_s'' \leq 0$ [39]. The choice of signs of the real and imaginary parts of $k_s$ and $\eta_s$ will be discussed below. As shown in Figure 1, a Cartesian coordinate system with the $(x, y, z)$ and the corresponding unit vectors $(\hat{x}, \hat{y}, \hat{z})$ is introduced such that the front face of the slab coincides with the $z = 0$ plane and that its back face coincides with the $z = d$ plane.

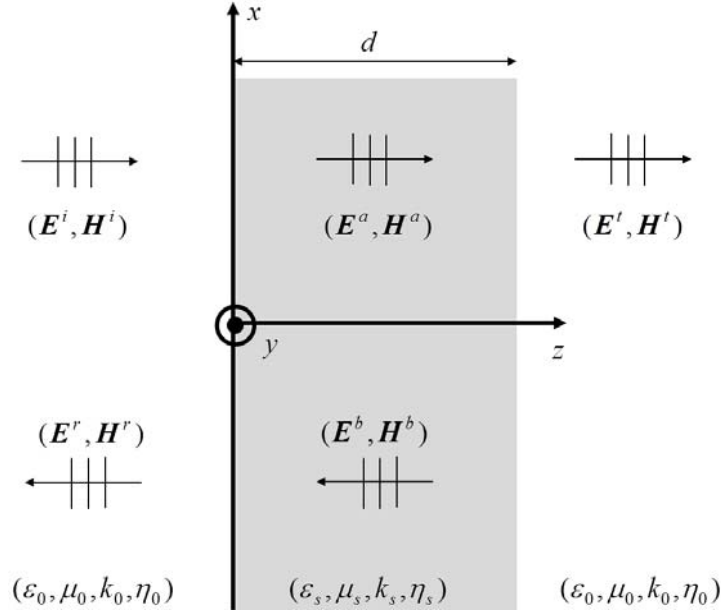

**Figure 1:** Normal uniform plane wave incidence on a homogeneous slab: the various fields, geometry and electromagnetic properties of the configuration.

*2.2 Field solutions*

The incident electric, $\boldsymbol{E}^i$, and magnetic, $\boldsymbol{H}^i$, fields can be expressed as

$$\boldsymbol{E}^i(z) = \hat{x} E^i e^{-jk_0 z} \text{ and } \boldsymbol{H}^i(z) = \hat{y} \frac{E^i}{\eta_0} e^{-jk_0 z}, \tag{1}$$

while the reflected electric, $\boldsymbol{E}^r$, and magnetic, $\boldsymbol{H}^r$, fields can be expressed as

$$\boldsymbol{E}^r(z) = \hat{x} E^r e^{jk_0 z} \text{ and } \boldsymbol{H}^r(z) = \hat{y} \frac{E^r}{\eta_0} e^{jk_0 z}. \tag{2}$$

The electric and magnetic fields within the slab can be represented by linear combinations of the fields

$$\boldsymbol{E}^a(z) = \hat{x} E^a e^{-jk_s z} \text{ and } \boldsymbol{H}^a(z) = \hat{y} \frac{E^a}{\eta_s} e^{-jk_s z}, \tag{3}$$

and

$$\boldsymbol{E}^b(z) = \hat{x} E^b e^{jk_s z} \text{ and } \boldsymbol{H}^b(z) = \hat{y}\frac{E^b}{\eta_s} e^{jk_s z}. \tag{4}$$

In the region $z > d$, there exist transmitted electric, $\boldsymbol{E}^t$, and magnetic, $\boldsymbol{H}^t$, fields

$$\boldsymbol{E}^t(z) = \hat{x} E^t e^{-jk_0 z} \text{ and } \boldsymbol{H}^t(z) = \hat{y}\frac{E^t}{\eta_0} e^{-jk_0 z}. \tag{5}$$

The quantity $E^i$ in (1) is the known amplitude of the incident electric field; $E^r$ in (2) is the amplitude of the reflected field; $E^a$ and $E^b$ in (3) and (4) are the amplitudes of the two electric fields within the slab; while $E^t$ in (5) is the amplitude of the transmitted field. The unknown amplitudes $E^r$, $E^a$, $E^b$, and $E^t$ follow from the enforcement of the boundary conditions at the $z=0$ and $z=d$ planes, and the result can be expressed as[1]

$$E^r = E_1^i \frac{(1-Z^2)(\eta_s^2 - \eta_0^2)}{D}, \tag{6a}$$

$$E^a = E^i \frac{2\eta_s(\eta_0 + \eta_s)}{D}, \tag{6b}$$

$$E^b = E^i \frac{2Z^2 \eta_s(\eta_0 - \eta_s)}{D}, \tag{6c}$$

$$E^t = E^i \frac{4\eta_0 \eta_s Z e^{jk_0 d}}{D}, \tag{6d}$$

with

$$D = (\eta_s + \eta_0)^2 - (\eta_s - \eta_0)^2 Z^2, \tag{7}$$

$$Z = e^{-jk_s d}. \tag{8}$$

### 2.3 Signs of $k_s$ and $\eta_s$

With the field solution (6-8) in place, the signs of the wave number $k_s$ and intrinsic impedance $\eta_s$ are now examined and their influence on the solution is discussed. Through Maxwell's curl equations, the parameters $k_s, \eta_s, \varepsilon_s$ and $\mu_s$ are related as

$$\eta_s = \frac{\omega \mu_s}{k_s} = \frac{k_s}{\omega \varepsilon_s}, \tag{9a}$$

$$k_s^2 = \omega^2 \varepsilon_s \mu_s, \tag{9b}$$

Thus, with a proper choice of the complex square root, it follows that

$$\mu_s = \sqrt{\frac{\mu_s}{\varepsilon_s}}, \tag{9c}$$

$$k_s = \omega\sqrt{\varepsilon_s \mu_s}. \tag{9d}$$

---

[1] As will be clear from Section 2.3, the unknown amplitudes in (6) can take on different, but equivalent, forms, this leading to different, but equivalent, expressions for the S-parameters. Those presented in (6) yield the S-parameters in [21], *cf.* Section 2.4.

From (9a) it follows that a change of sign of $k_s$ implies a simultaneous change of sign of $\eta_s$, if the permittivity or permeability are specified. As to the field behavior upon the change of signs of $k_s$ and $\eta_s$, one notes from (6) that

$$E^r(k_s \to -k_s; \eta_s \to -\eta_s) \to E^r, \tag{10a}$$

$$E^a(k_s \to -k_s; \eta_s \to -\eta_s) \to E^b, \tag{10b}$$

$$E^b(k_s \to -k_s; \eta_s \to -\eta_s) \to E^a, \tag{10c}$$

$$E^t(k_s \to -k_s; \eta_s \to -\eta_s) \to E^t, \tag{10d}$$

since

$$Z(k_s \to -k_s; \eta_s \to -\eta_s)_s \to 1/Z, \tag{11}$$

where $Z$ is defined in (8). Thus, a change in sign of $k_s$, and thus of $\eta_s$, has no effect on the field solution of the configuration shown in Figure 1. The two fields inside the slab are seen to merely switch their roles with no effect on the total field inside the slab. Moreover, the fields outside remain exactly the same.

Both signs of the wave number $k_s$ and, hence, the intrinsic impedance $\eta_s$ lead to the same field solution. Neither of the signs can thus be discarded by physical arguments, such as passivity. Again, the simple reason is, of course, that the wave number and intrinsic impedance – unlike the permittivity and permeability – are not fundamental quantities in Maxwell's equations, but are derived quantities introduced for convenience. When introduced they can be defined with one sign or the other, as long as either definition is followed stringently.

*2.4 Scattering parameters*

The reflection and transmission properties of the homogeneous slab in Figure 1 are fully accounted for through its reflection, $S_{11}$, and transmission, $S_{21}$, coefficients, which are the two S-parameters of importance. Defining $S_{11}$ and $S_{21}$ at the reference planes located at $z = 0$ and $z = d$, respectively, we can derive from (6)-(8) the following S-parameter expressions, which are equivalent to those derived originally in the Nicolson-Ross-Weir method [21, 22]:

$$S_{11} \equiv \frac{\hat{x} \cdot \boldsymbol{E}^r\big|_{z=0}}{\hat{x} \cdot \boldsymbol{E}^i\big|_{z=0}} = \frac{E^r}{E^i} = \frac{(1-Z^2)(\eta_s^2 - \eta_0^2)}{(\eta_s + \eta_0)^2 - (\eta_s - \eta_0)^2 Z^2}, \tag{13a}$$

$$S_{21} \equiv \frac{\hat{x} \cdot \boldsymbol{E}^t\big|_{z=d}}{\hat{x} \cdot \boldsymbol{E}^i\big|_{z=0}} = \frac{E^t e^{-jk_0 d}}{E^i} = \frac{4\eta_0 \eta_s Z}{(\eta_s + \eta_0)^2 - (\eta_s - \eta_0)^2 Z^2}. \tag{13b}$$

The expressions in (13) are, moreover, identical to those derived in [21]. For later use, we note that others, see, *e.g.*, [23], have derived equivalent expressions for the S-parameters by first introducing a transfer matrix for the configuration in Figure 1, and subsequently converting it to the scattering matrix. With the time convention and symbols adopted in the present manuscript, the results of [23], which likewise are equivalent to the original results in [24, 25], read

$$S_{21} = \frac{1}{\cos(k_s d) + \frac{j}{2}(\frac{\eta_s}{\eta_0} + \frac{\eta_0}{\eta_s})\sin(k_s d)}, \qquad (14a)$$

$$S_{11} = S_{21} \frac{j}{2} \sin(k_s d)(\frac{\eta_s}{\eta_0} - \frac{\eta_0}{\eta_s}). \qquad (14b)$$

In the original Nicolson-Ross-Weir method [24], [25], the S-parameters are expressed as

$$S_{11} = \frac{(1-Z^2)\Gamma}{1-\Gamma^2 Z^2}, \qquad (15a)$$

$$S_{12} = \frac{(1-\Gamma^2)Z}{1-\Gamma^2 Z^2}, \qquad (15b)$$

where

$$\Gamma = \frac{\eta_s - \eta_0}{\eta_s + \eta_0},$$
(15c)

is the reflection coefficient for a half-space and the quantity $Z$ is given by (8). We reiterate that the S-parameter expressions (13), (14), and (15) are equivalent.

*2.5 Ambiguities for permittivity and permeability*

There are multiple sets of the slab permittivity $\varepsilon_s$ and permeability $\mu_s$ that give the same S-parameters, $S_{11}$ and $S_{21}$, and the S-parameter extraction method thus becomes ambiguous. In Section 3, the ambiguities will be described mathematically using an inversion of (13). Here, these ambiguities can be readily illustrated through the forward problem. Obviously, from the result in (13), it follows that the S-parameters will remain unchanged if $\eta_s$ and $Z$ do not change. Consequently, in order for two different sets of slab material parameters, ($\varepsilon_{s1}, \mu_{s1}$) and ($\varepsilon_{s2}, \mu_{s2}$), to yield the same S-parameters, it follows from (9c) and (9d) that

$$\frac{\mu_{s1}}{\varepsilon_{s1}} = \frac{\mu_{s2}}{\varepsilon_{s2}}, \qquad (16)$$

for $\eta_s$ to remain unchanged, and

$$d\omega\sqrt{\varepsilon_{s1}\mu_{s1}} = d\omega\sqrt{\varepsilon_{s2}\mu_{s2}} + 2p\pi, \quad p \in \mathbf{Z} \qquad (17)$$

for $Z$ to remain unchanged. In (17), the symbol $\mathbf{Z}$ denotes the set of all integers. The solution to (16) and (17) is given by

$$\varepsilon_{s2} = \varepsilon_{s1} X, \qquad (18)$$

$$\mu_{s2} = \mu_{s1} X, \qquad (19)$$

with

$$X = \sqrt{1 + \frac{1}{\varepsilon_{s1}\mu_{s1}}\left(\frac{p\pi}{d\omega} - \sqrt{\varepsilon_{s1}\mu_{s1}}\right)\frac{4p\pi}{d\omega}} \quad , \quad p \in \mathbf{Z} \tag{20}$$

where the branch of the complex square root in (20) is chosen such that $\mathrm{Im}(\varepsilon_{s2}) \leq 0$ and $\mathrm{Im}(\mu_{s2}) \leq 0$. It thus follows from (19)-(20) that for a given frequency and length of the slab, there are infinitely many slab material parameters that will result in the same S-parameters.

In order to illustrate these matters, consider the example in which $\varepsilon_{s1} = \varepsilon_0(2 - j0.5)$ and $\mu_{s1} = \mu_0(3 - j0.5)$ when the thickness of the slab is $d = \lambda_0/2\pi$, where $\lambda_0$ is the free-space wavelength. Since $\lambda_0 = c_0/f$ and $\omega = 2\pi f = 2\pi c_0/\lambda_0$, where $c_0 = 1/\sqrt{\varepsilon_0\mu_0}$ is the speed of light in free-space, and $f$ is the operating frequency, the present choice of the slab thickness makes the product $d\omega$ in (20) equal to $c_0 = 1/\sqrt{\varepsilon_0\mu_0}$. Figure 2 illustrates the different sets of the permittivity $\varepsilon_{s2}$ and permeability $\mu_{s2}$ obtained for $p = -3, -2, -1, 0, +1, +2, +3$ that give the same S-parameters as $\varepsilon_{s1}$ and $\mu_{s1}$.

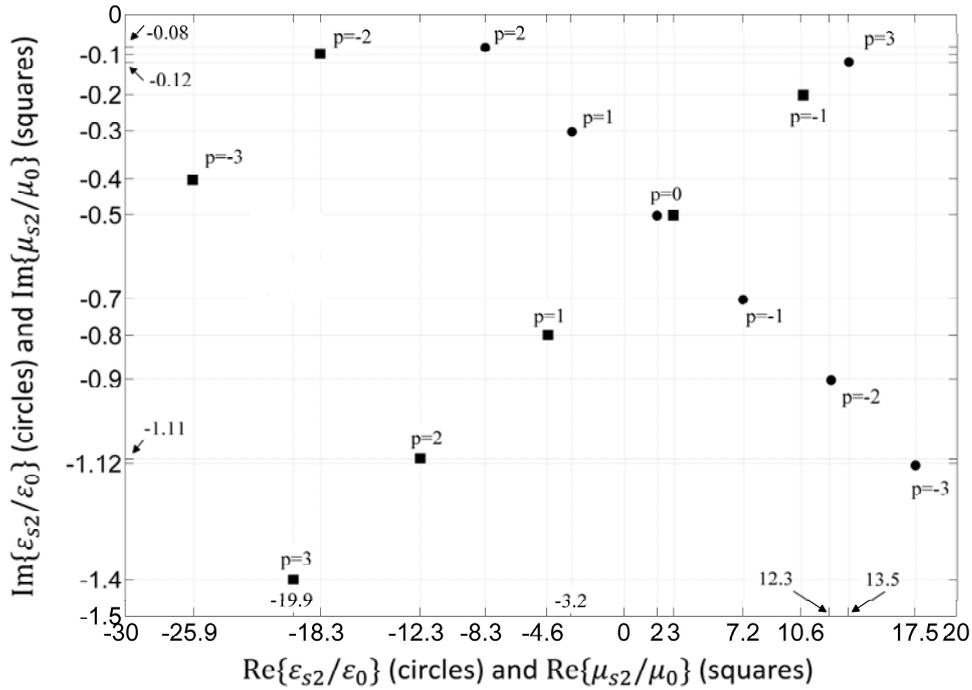

**Figure 2:** The different sets of the permittivity $\varepsilon_{s2}$ and permeability $\mu_{s2}$ that give the same S-parameters as the slab material parameters $\varepsilon_{s1} = \varepsilon_0(2 - j0.5)$ and $\mu_{s1} = \mu_0(3 - j0.5)$. The depicted $\varepsilon_{s2}$ and $\mu_{s2}$ have been obtained for $p = -3, -2, -1, 0, +1, +2, +3$ in (18)-(20) and the slab thickness was selected to $d = \lambda_0/2\pi$. The quantity $\varepsilon_{s2}/\varepsilon_0$ is represented by circles, while $\mu_{s2}/\mu_0$ is represented by squares. See the main text for further explanations.

The horizontal and vertical axes in the figure are the real and imaginary parts, respectively, of the quantity $\varepsilon_{s2}/\varepsilon_0$, represented by circles, and $\mu_{s2}/\mu_0$, represented by squares. For $p=0$, the parameters $\varepsilon_{s2}$ and $\mu_{s2}$ are, of course, equal to $\varepsilon_{s1}$ and $\mu_{s1}$; this is also easily seen from (18)-(20) since $X=1$ in (20) for $p=0$. However, for other values of $p$ it is seen that different sets of $\varepsilon_{s2}$ and $\mu_{s2}$ result which give the same S-parameters as $\varepsilon_{s1}$ and $\mu_{s1}$. In particular, it is interesting to note that sets of $\varepsilon_{s2}$ and $\mu_{s2}$ exist for which their real parts are both positive ($p=-1$), both negative ($p=1, 2$), and with one being positive and the other being negative ($p=-2,-3$ and $+3$). Therefore, due to these different possibilities that occur for various values of $p$, the effective material parameters for a given material can not be interpreted solely on the basis of its S-parameters as the ones which possesses, *e.g.*, negative real parts. Thus, great care needs to be exercised towards the correct selection of the value of $p$ in order to properly interpret the material parameters of a given MTM based on their extraction from the S-parameters. These matters are further illustrated with specific examples in Section 4.

**3. Inverse problem - extraction of material parameters**

The idea behind the S-parameter extraction technique is to solve (13) (or, equivalently, (14) or (15)) for $k_s$ and $\eta_s$, from which the permittivity and permeability of the slab can be determined through the following relations

$$\mu_s = \frac{k_s \eta_s}{\omega}, \qquad (21a)$$

$$\varepsilon_s = \frac{k_s}{\omega \eta_s}. \qquad (21b)$$

The following discussion is initialized by taking the S-parameter expressions (13) and solving for $\eta_s$ with the result

$$\eta_s = \pm \eta_0 \sqrt{\frac{(S_{11}+1)^2 - S_{21}^2}{(S_{11}-1)^2 - S_{21}^2}}. \qquad (22)$$

This expression has also been reported in [18], [20], [22], and [23]. With the impedance $\eta_s$ known, there are two formally different possibilities of obtaining $Z$ (as given in (8)) (and thus $k_s$) from the S-parameters in (13). One possibility, referred to as $Z_A$, follows by solving for $Z^2$ in the relation (13a) for $S_{11}$ and substituting this into the $S_{21}$ expression (13b) in order to arrive at

$$Z_A = \frac{S_{21}(\eta_s + \eta_0)}{(\eta_s + \eta_0) - S_{11}(\eta_s - \eta_0)}. \qquad (23a)$$

Another possibility, referred to as $Z_B$, follows by solving instead for $(\eta_s - \eta_0)^2$ in (13a) and substituting the result into (13b) to arrive at

$$Z_B = \frac{(\eta_s - \eta_0) - S_{11}(\eta_s + \eta_0)}{S_{21}(\eta_s - \eta_0)}. \qquad (23b)$$

With $Z$ (as given in (8)) determined from either (23a) or (23b), the expression for $k_s$ is found to be

$$k_s = \frac{j}{d}\log Z = \frac{1}{d}\left[-\arg Z + j\text{Log}|Z|\right] = \frac{1}{d}\left[-(\text{Arg}\,Z + 2p\pi) + j\text{Log}|Z|\right], \quad p \in \mathbf{Z} \qquad (24)$$

where log Z denotes the multiple-valued complex natural logarithm of Z, arg Z is the multiple-valued argument of Z, Arg Z is the principle branch of the argument of Z, and Log Z denotes the principal branch of the real part of the logarithm of Z. The term $2\pi p$ with $p \neq 0$ (where p is the branch index) defines branches of log Z other than the principal one and, thus, it gives the branch ambiguity in the real part of $k_s$. It is noted that there is no branch ambiguity in the imaginary part of $k_s$. The simple physical reason for this is that while the phase can only be measured with a $2\pi p$ ambiguity, the loss can be measured absolutely.

All the ingredients of the S-parameter extraction technique have now been obtained and the procedure is summarized in the following steps:

1. obtain the S-parameters, $S_{11}$ and $S_{21}$, of a slab with known thickness $d$,
2. determine the intrinsic impedance $\eta_s$ from (22),
3. determine the term Z from (23)
4. determine the wave number $k_s$ from (24), and finally
5. determine the material parameters $\varepsilon_s$ and $\mu_s$ from (21).

Having clarified the details of the S-parameter approach, we next demonstrate that both signs of the intrinsic impedance $\eta_s$ (20) and the accompanying signs of wave number $k_s$ (24) are equally valid since they lead to the same material parameters in (21).

It is easily shown that $Z_A = Z_B$, and moreover, that

$$Z_A(\eta_s \to -\eta_s) \to \frac{1}{Z_A}, \qquad (25)$$

which is thus the same as $1/Z_B$. Therefore, the expression for $Z_A$ (23a) should, under the change $\eta_s \to -\eta_s$, be replaced by $1/Z_A$. As a consequence,

$$k_s(\eta_s \to -\eta_s) \to \frac{j}{d}\log\left(\frac{1}{Z}\right) = -\frac{j}{d}\log Z = -k_s, \qquad (26)$$

which is also in agreement with (9). Thus, changing the sign of $\eta_s$ simultaneously changes the sign of $k_s$, thereby having no effect on the sign of either their product (for the determination of $\mu_s$ in (21a)) or ratio (for the determination of $\varepsilon_s$ in (21b)), and thus on the resulting field solution of the configuration depicted in Figure 1. These observations are in agreement with our discussion in Section 2.3.

Identical conclusions are reached if the S-parameters given by (14), which have been derived in [23], are taken in the outset. In [23] the expression for $\eta_s$, identical to the one in (22), has been derived, whereas the wave number $k_s$, with the appropriate modifications of the time convention and symbols, satisfies the expression

$$\cos(k_s d) = \frac{1 - S_{11}^2 + S_{21}^2}{2S_{21}}. \qquad (27a)$$

The presence of the branch ambiguity in determining the wave number $k_s$ from (27a) can not be questioned, owing to the different branches of the inverse cosine function. However, we note that according to (14b), $k_s$ furthermore needs to satisfy the relation

$$\sin(k_s d) = \frac{2}{j} \frac{S_{11}}{S_{21}} \left( \frac{\eta_s}{\eta_0} - \frac{\eta_0}{\eta_s} \right)^{-1}, \qquad (27b)$$

from which it again follows that a change of sign of $\eta_s$ changes the sign of $k_s$, i.e., the two signs are dependent. Consequently, either sign of $k_s$ in (27a) can be used without affecting the extracted material parameters in (21).

In summary, the S-parameter method holds no ambiguity in the sign of the wave number and intrinsic impedance. A change of sign in one of these parameters leads to a simultaneous change of sign of the other parameter, i.e., their signs are dependent and both signs lead to the same permittivity and permeability. Thus, neither of the signs can be discarded by physical arguments such as passivity. This is in contrast to some previous reports [18]-[23] where passivity was used to select the proper signs of the intrinsic impedance $\eta_s$ in (22) and the associated sign of wave number $k_s$ in (24) (or (27)) by requiring that real part of the intrinsic impedance, $\text{Re}(\eta_s)$ and the imaginary part of the wave number $\text{Im}(k_s)$ (or equivalently the imaginary refractive index $n_s$, $\text{Im}(n_s)$), satisfy $\text{Re}(\eta_s) \geq 0$, $\text{Im}(k_s) \leq 0$ ($\text{Im}(n_s) \leq 0$).

The S-parameter extraction method therefore only contains the branch ambiguity, which appears in the real part of the wave number as branches of the complex logarithm (in (24)) or the complex inverse cosine (in (27)).

## 4. Illustration of the branch ambiguity

The S-parameter method has been applied successfully to retrieve the material parameters of a number of MTM samples thereby demonstrating its applicability for MTM characterization [7], [18]-[23], [26]. The purpose of the present section is to illustrate the significance of the choice of the branch of the complex logarithm in (24). Specifically, it is shown that rather different material parameters, can result as different branches in (24) are selected in the extraction process. This is done through simulations of first a slab with constant and known material parameters and second of a specific MTM design reported in [38].

### 4.1 A slab with constant material parameters

We consider a slab consisting of a simple, lossless and non-dispersive material with $\varepsilon_s / \varepsilon_0 = 2$ and $\mu_s / \mu_0 = 1$ and $d = 10$ cm. The S-parameters of the slab are calculated by (13) in the frequency range $f \in [0, 8.5]$ GHz; thus, the range of the electrical thickness of the slab $k_s d \in [0, 8\pi]$. Upon inversion of the S-parameters, the extracted material parameters are easily found; they are shown in Figure 3.

The results in Figure 3 show that the correct material parameters, $\varepsilon_s / \varepsilon_0 = 2$ and $\mu_s / \mu_0 = 1$, can be extracted by selecting the proper value of $p$. However, they also show that the choice is dependent on the frequency. It is seen that as one follows, for increasing $k_s d$ values, a curve for any given $p$ value, a discontinuity in the material parameters occurs at some point. On the other hand, the $p+1$ curve then takes over to form a continuous result. The correct $p$ curve for vanishing $k_s d$ is, of course, $p = 0$, but changes from 0 to 1 at $k_s d / 2\pi = 0.5$ and from 1 to 2 at $k_s d / 2\pi = 1.5$ and so on. One can select the proper value of $p$ by obtaining the low frequency values of the material parameters and then change the value to ensure their continuity. However, it is doubtful whether this process can be automated for general structures and used in systematic design procedures, e.g., [40], since branch crossings and other special features may obscure the choice of correct branches. It is interesting to observe that the discontinuities in Figure 3 are smaller when the absolute value of $p$ is larger. Due to these discontinuities, the extracted material parameters would obviously be wrong if the $p$ value were chosen improperly. It is found that, depending on the chosen $p$ value, an actual material with positive material parameters can be mistakenly interpreted as one having negative material parameters and vice versa. Thus, the branch ambiguity and, therefore, the selection of the proper value of $p$, must carefully be addressed in order to provide meaningful material characterization using the S-parameter measurements.

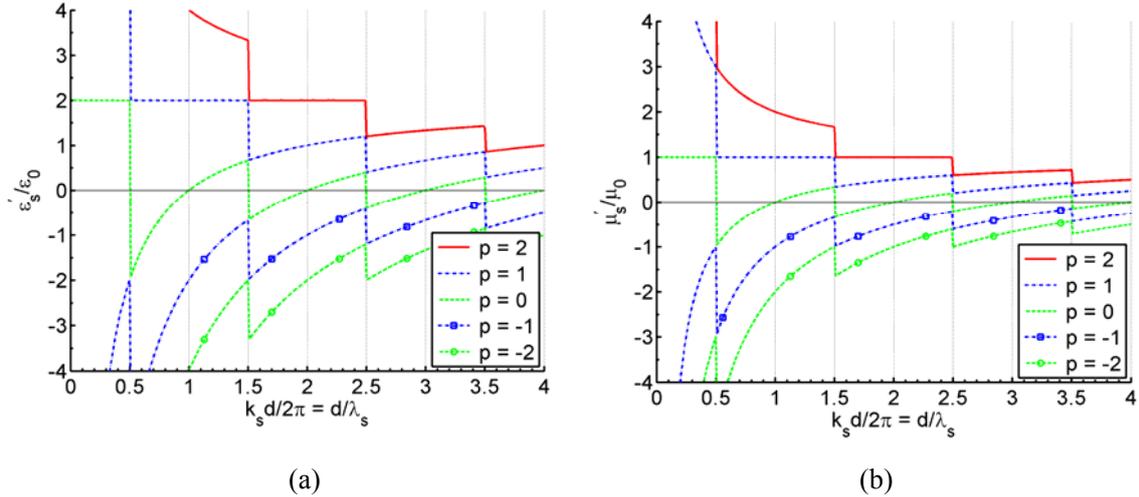

(a)          (b)

Figure 3: Extracted material parameters for a simple, lossless, and non-dispersive slab with $\varepsilon_s / \varepsilon_0 = 2$ and $\mu_s / \mu_0 = 1$ and $d = 10$ cm. (a) $\varepsilon_s / \varepsilon_0$, and (b) $\mu_s / \mu_0$.

*4.2 Specific MTM design – S-shaped resonators*

In [38] it was shown how a material with negative real parts of the permittivity and permeability, a so-called double negative (DNG) material, can be synthesized over a wide range of frequencies, i.e., from 15 GHz to 20 GHz, by using S-shaped resonators, see also [38, Figure 2(b)]. In the following we have replicated these results in ANSOFT HFSS software for a slab thickness of 4 mm in order to illustrate the significance of the branch ambiguity on a realistic MTM design. The thickness of the conducing S-shaped structure was not specified in [38]; inhere perfectly electrically conducting S-shaped structures with a thickness of $35\mu m$ were used in our simulations to replicate the results from [38, Figure 2(b)]. The extracted real parts of the permittivity and permeability are shown in Figure 4 for the two branch indexes: $p = 0$ and $p = 1$.

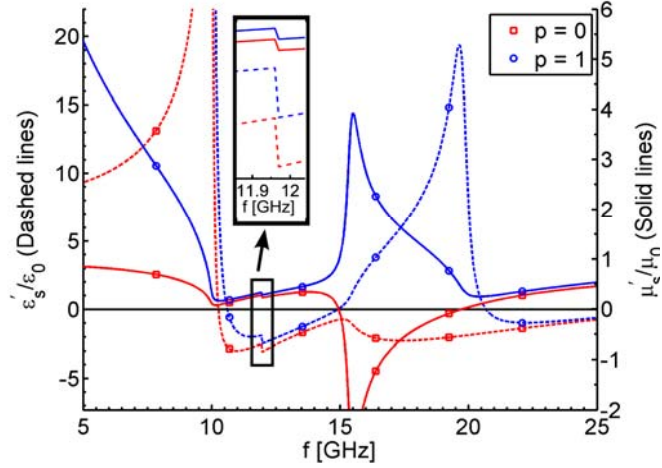

**Figure 4:** The extracted real parts of the permittivity and permeability for the S-shaped unit cell. The results are shown for the two branches with indexes $p=0$ and $p=1$.

Comparing the $p=0$ results in Figure 4 with those in [38, Figure 2(b)], one observes a good qualitative agreement between the results in the frequency range of interest. This suggests that the value $p=0$ was selected in [38]. A negative permittivity was observed in [38] in the frequency interval from 12 to 25 GHz, while a negative permeability was observed in the frequency interval from 15 to 20 GHz; thus a DNG region from 15 GHz to 20 GHz was reported in [38, Figure 2(b)]. In our extraction with $p=0$, a negative permittivity is observed in the range from 10.4 GHz to 25 GHz and a negative permeability is observed in the range from 15 GHz to 20 GHz. However, as indicated in the inset in Figure 4, a discontinuity of all results with the branch index $p=0$ occurs around 11.95 GHz. At that frequency, a change of $p$ from 0 to 1 must occur in order to obtain continuous results for the effective material parameters. By performing such a change from $p=0$ to $p=1$ at 11.95 GHz it is seen that the permittivity is actually positive in the region from 15 GHz to 20.5 GHz region, whereas it was reported negative in this region in [38, Figure 2(b)] of negative permittivity reported in [38] is actually positive. This was also the range where the negative permeability was seen in [38]. Thus, if both $p=0$ and $p=1$ are used (which must be the case to ensure the continuity of the effective material parameters), the S-shaped resonator from [38] is interpreted as a negative permittivity material only (now in dramatically decreased frequency regions from 10-15 GHz and again from around 21-25 GHz), while it is found to be a DNG material if only the $p=0$ branch is used throughout the frequency range of interest. In view of these findings, we thus believe that the result in [38, Figure 2(b)] does not represent the correct material parameters of the S-resonator DNG material studied in [38]. It is therefore clear that different interpretations of materials in terms of their material parameters obtained by the S-parameter method may occur unless great care is devoted to the branch ambiguity and the selection of a proper value of $p$.

**5. Solution to branch ambiguity**

A review of the potential approaches for solving the branch ambiguity in (24) (or (27)) is next presented and their applicability to MTMs is discussed. There are essentially two categories of slabs to address: 1) electrically thin slabs, and 2) electrically thick slabs.

*5.1 Electrically thin slabs*

When the material slab depicted in Figure 1 is electrically thin, *i.e.*, when the wavelength, $\lambda_s$, inside the slab satisfies $\lambda_s > 2d$, with $d$ being the thickness of the slab, the S-parameter extraction method has no ambiguity since the unique solution results by choosing the principal branch of the complex logarithm in (24) for which $p=0$. This follows at once by rewriting (8) in the following manner

$$Z = e^{-jk_s d} = e^{-jk_s' d} e^{k_s'' d}, \qquad (28)$$

from which $\arg Z = -k_s' d = \mathrm{Arg}\, Z + 2p\pi$, $p = 0, \pm 1, \pm 2, \ldots$ Since $\lambda_s > 2d$, which corresponds to $|k_s'| d < \pi$, then Arg $Z$ is limited to the interval $]-\pi, \pi]$ and one can set $p=0$ in (24). Thus, the material parameters can be extracted unambiguously for electrically thin slabs. However, this approach is based on an *a priori* knowledge of $k_s'$ and thus of the wavelength $\lambda_s$, which might not be available in practice.

It is worth mentioning that electrically thin slabs can be modeled as metasurfaces (metafilms). As shown in, e.g., [41, 42], the latter can be characterized unambiguously using the generalized sheet transition conditions.

*5.2 Electrically thick samples*

For electrically thick slabs, the principal branch can not, in general, be chosen. Several procedures for proper branch selection exist.

*5.2.1 Two different lengths*

An established method of eliminating the branch ambiguity is to obtain the wave number in (24) for two different lengths, $d_1$ and $d_2 > d_1$, of the same slab, and then determine the branches which give the real part of the wave number, $k_{s,1} = k_{s,1}' + jk_{s,1}''$, for the slab of thickness $d_1$, equal to the real part of the wave number, $k_{s,2} = k_{s,2}' + jk_{s,2}''$, for the slab of thickness $d_2$ (recall that there is no ambiguity in the imaginary part of wave number). This method was used in [43] for the characterization of lossy dielectric materials. In analogy to (24), the two wave numbers are given by

$$k_{s,1} = \frac{j}{d_1} \log Z_1 = \frac{1}{d_1}\left[-\arg Z_1 + j\mathrm{Log}|Z_1|\right] = \frac{1}{d_1}\left[-(\mathrm{Arg}\, Z_1 + 2p_1\pi) + j\mathrm{Log}|Z_1|\right], \qquad (29a)$$

$$k_{s,2} = \frac{j}{d_2} \log Z_2 = \frac{1}{d_2}\left[-\arg Z_2 + j\mathrm{Log}|Z_2|\right] = \frac{1}{d_2}\left[-(\mathrm{Arg}\, Z_2 + 2p_2\pi) + j\mathrm{Log}|Z_2|\right], \qquad (29b)$$

where $p_1 \in \mathbf{Z}$, $p_2 \in \mathbf{Z}$, and $Z_i = \exp(-jk_{s,i} d_i)$, $i$=1 and 2.

If the angle measured clockwise in complex plane from $Z_1$ to $Z_2$ is less than $\pi$, which requires $2(d_2 - d_1) < \lambda_s$, i.e., an electrically thin difference between the two slab widths $d_1$ and $d_2$, then $p_1 = p_2 \equiv p$, in which case the identity of the real parts of $k_{s,1}$ and $k_{s,2}$ in (29) gives

$$p = \frac{d_2 \operatorname{Arg} Z_1 - d_1 \operatorname{Arg} Z_2}{2\pi(d_1 - d_2)}. \tag{30}$$

This method, however, requires an *a priori* knowledge of the wavelength $\lambda_s$, and thus $k_s'$. Therefore, the same practical problem as for an electrically thin slabs treated in Section 5.1 exists.

For an electrically thick difference between the two slab widths $d_1$ and $d_2$, one has $p_1 \neq p_2$. In this case the identity of the real parts of $k_{s,1}'$ and $k_{s,2}'$ in (29) gives

$$p_2 d_1 - p_1 d_2 = \frac{d_2 \operatorname{Arg} Z_1 - d_1 \operatorname{Arg} Z_2}{2\pi}. \tag{31}$$

Since $p_1$ and $p_2$ are integers, a possible approach is to try out all integers in some reasonable interval and determine those satisfying (31). In order to investigate whether a given set of $p_1$ and $p_2$ is unique, it is instructive to try out two different sets of integers and check if they lead to the same result. Expressing the second set as $p_1 + m$ and $p_2 + n$, one has

$$p_2 d_1 - p_1 d_2 = (p_2 + n) d_1 - (p_1 + m) d_2 \Leftrightarrow 0 = -(md_2 - nd_1) \Leftrightarrow \frac{m}{n} = \frac{d_1}{d_2}. \tag{32}$$

The result in (32) shows that if the ratio of $m$ and $n$ equals the ratio of $d_1$ and $d_2$, these integers can be added to $p_1$ and $p_2$, respectively, without altering the result. If the ratio $d_1/d_2$ is a rational number, the quantity $m/n$ exists and the set $(p_1, p_2)$ is not unique. If the ratio $d_1/d_2$ is an irrational number, the set $(p_1, p_2)$ is unique.

*5.2.2 Multiple frequencies*

For weakly dispersive media, a common approach for elimination of the branch ambiguity is the so-called group delay method. For weakly dispersive media, the phase velocity, $v_p \cong \omega/k_s'$, is an almost linear function of the angular frequency $\omega$, and the group velocity, $v_g = \partial\omega/\partial k_s'$ is nearly equal to $v_p$. The group delay, $\tau_g$, across the sample in Figure 1, reads

$$\tau_g = \frac{d}{v_g} = d \frac{\partial k_s'}{\partial \omega}, \tag{33}$$

while the phase delay, $\tau_p$, is given by

$$\tau_p = \frac{d}{v_p} \cong d \frac{k_s'}{\omega}. \tag{34}$$

It is now possible to choose the correct solution for $k_s$, and thus select the correct value of $p$, by imposing the requirement that the absolute difference between the group and phase delays in (33) and (34) should be minimal. It is obvious that measurements need to be taken at two or more frequencies in order to calculate the derivative in (33). As noted in the beginning of this section, the group-delay method applies well for weakly dispersive media. As such, it is not applicable for highly dispersive or lossy MTM samples.

The expressions in (33) and (34) might leave the impression that both the group delay and phase delay depend on the branch index $p$. However, the differentiation of the wave number in (24) explicitly shows that the former is independent of the branch index $p$. To also illustrate this in the context of the refractive index $n_s(\omega) = n_s'(\omega) + jn_s''(\omega)$, we note that for a general homogenous medium the dispersion relation can now be expressed as $k_s(\omega) = \omega n_s(\omega)/c_0$, so that (33) now becomes

$$\tau_g = \frac{d}{c_0}\left(n_s' + \omega \frac{\partial n_s'}{\partial \omega}\right). \tag{35}$$

Obviously, the second term accounts for the dispersion in the medium. The different branches can now be addressed. Since $n_s'(\omega) = n_{s,p=0}'(\omega) + [2p\pi/(k_0(\omega)d)]$ [28], it follows by explicit differentiation that the two terms in (35) will serve to balance the apparent dependence on the branch index. We may conveniently rewrite (35) as

$$\tau_g = \tau_p + k_0(\omega)d \frac{\partial n_s'}{\partial \omega}, \tag{36}$$

which perfectly illustrates that $\tau_g \approx \tau_p$ for a weakly dispersive medium where $\partial n_s'/\partial \omega \approx 0$. In turn, this implies that

$$p \approx \frac{k_0(\omega)d}{2\pi}\omega \frac{\partial n_{s,p=0}'}{\partial \omega}, \tag{37}$$

and we inherently find the branch $p = 0$ to stay consistent with the weak dispersion assumed in the first place.

*5.2.3 Kramers-Krönig relations*

To circumvent the problems that some of the common methods described above encounter with MTMs, another method has recently been proposed in [44, 45]. The well-known Kramers-Krönig relationships, which link the real and imaginary parts of the wave number, were used to select the correct value of the branch index $p$. The main idea exploits the fact that the imaginary part of the wave number is determined uniquely, and it can therefore, via the Kramers-Krönig relationship, be used to reconstruct the correct value of the real part of the wave number. Thus, it can serve as a guideline of finding the correct, unambiguous, value of $p$. Unfortunately, since the integrals involved in the Kramers-Krönig relationship are infinite (in the angular frequency), it is not clear how to estimate errors occurring when truncating the integrals. Moreover, as shown in [46], spatial dispersion effects, usually not taken into account in the basic retrieval schemes, can introduce artifacts into the data which will invalidate the Kramers-Krönig outcomes.

# 6. Summary and conclusions

In this work, the S-parameter extraction method was reviewed with the purpose of clarifying its ambiguity issues. This was accomplished through an investigation of a problem of a normally incident uniform plane wave on a planar slab located in free space.

The S-parameter method facilitates the extraction of the permittivity and permeability from measured, or otherwise known, S-parameters and gives the wave number and intrinsic impedance of the sample as intermediate steps. Using the forward problem we have demonstrated that the wave number and intrinsic impedance of the slab can be introduced with either sign without changing the physics of the problem. A change of the sign of the wave number was found to lead to a simultaneous change of the sign of the intrinsic impedance, and vice versa; thereby ensuring that either of the signs of these parameters can be associated with the same permittivity and permeability of the slab. A number of different – but equivalent – expressions for the S-parameters were reviewed. Additionally, different sets of permittivity and permeability values that lead to the same S-parameters were identified and illustrated with a specific example. The example clearly showed that same S-parameters can be obtained by permittivities and permeabilities whose real parts can be of either sign. These are the material parameters that constitute the ambiguous solutions in the S-parameter extraction method. They amply demonstrate that a material can mistakenly be considered as having e.g., negative permittivity and permeability, if the S-parameter extraction method is employed improperly.

The expressions for the S-parameters established in the forward problem were subsequently used in the inverse problem to provide relations for determining the permittivity and permeability of the slab. Based on the inverse problem expressions, we have demonstrated that both signs of the wave number and intrinsic impedance that follow from the inversion of the S-parameter expressions are equally valid and that the only ambiguity is the branch ambiguity related to the branch of the complex logarithm (or complex inverse cosine) which appears in the real part of the wave number. This demonstration was accomplished through a number of different – but equivalent – and rather well-known S-parameter expressions. The severity and the consequences of the branch ambiguity were illustrated through simulations of a slab with known constant material parameters and of an S-shaped resonator-based MTM design [38]. The former example showed that the S-parameter extraction method can be used to extract the correct material parameters if the proper branch of the complex logarithm is selected. Starting from the fundamental branch, the shift to the next branch was made at the frequency at which discontinuities in the material parameters arose. This shift ensured the continuous and thus correct material parameters. Similar shifts to even higher order branches were undertaken at the discontinuities of material parameters occurring at higher frequencies, with the overall result of extracting the correct and continuous material parameters. Applying the described procedure to a well-known S-shaped resonator – based MTM, we illustrated that such a structure exhibits the DNG properties, such as those reported in [38], only if the fundamental branch of the complex logarithm is selected throughout the entire frequency range. Moreover, we showed that by performing the shift of branch from the fundamental to the next higher order branch, in order to ensure the continuous material parameters, the S-shaped resonator structure was found to possess a negative permittivity in a dramatically reduced frequency range compared to that reported in [38] and no negative permeability at all. In view of these findings, we thus believe that the result in [38, Figure 2(b)] does not represent the correct material parameters of the S-resonator material studied in [38]. This demonstrates that different (and incorrect) interpretations of materials in terms of their effective material parameters obtained by the S-parameter extraction method may occur unless great care is exercised to the selection of a proper value of the branch of the complex logarithm (or the complex inverse cosine). Following these examples, a review of the potential approaches for solving the branch ambiguity of the S-parameter extraction method was presented. These included methods treating electrically thin and thick samples. For the latter, the method employing two samples of a different thickness, as well as those relaying on multiple frequencies and Kramers-Krönig relations, were reviewed. In all cases the challenges of applying them with success to MTMs were discussed. Among the reviewed methods for solving the branch ambiguity problem, the one employing two different lengths of the sample is the only one not requiring an *a priori* knowledge about wavelength and dispersive properties of the material, nor does it require measurements over very wide frequency ranges.

In conclusion, we recapitulate that the S-parameter extraction method only holds one ambiguity. This is the branch ambiguity of the complex logarithm (or complex inverse cosine) and it appears in the real part of the wave number, leading to ambiguous extracted permittivity and permeability. In contrast to some previous reports [18]-[23] we find that there is no sign ambiguity for the wave number and intrinsic impedance since both signs lead to the same permittivity and permeability and neither can thus be discarded by physical arguments such as passivity. The simple reason is that the wave number and intrinsic impedance – unlike the permittivity and permeability – are not fundamental quantities in Maxwell's equations or the associated constitutive relations, but are derived quantities that are introduced for convenience. When introduced, they can be defined with one sign or the other and as long as either definition is followed stringently.

**Acknowledgements**

This work is supported in part by the Danish Research Council for Technology and Production Sciences within the TopAnt project (www.topant.dtu.dk). The work by RWZ was supported in part by DARPA Contract number HR0011-05-C-0068 and by ONR Contract number H940030920902. Since the original submission of this manuscript, a relevant work [47] has been published considering an approach to the branch ambiguity problem.